\documentclass[prb,aps,twocolumn]{revtex4}
\usepackage{graphicx,color,latexsym}
\usepackage{dcolumn}
\usepackage{amsmath,amssymb,epsf,bm}
\begin{document}
\title{Supercurrent vortices and Majorana zero modes induced by an inplane Zeeman field on the surface of a three-dimensional topological insulator}
\author{A.~G. Mal'shukov}
\affiliation{Institute of Spectroscopy, Russian Academy of Sciences, Troitsk, Moscow, 108840, Russia}
\affiliation{Moscow Institute of Physics and Technology, Institutsky per.9, Dolgoprudny, 141700 Russia}
\affiliation{National Research University Higher School of Economics, Myasnitskaya str. 20, Moscow, 101000 Russia}
\begin{abstract}
A nonuniform in-plane Zeeman field can induce spontaneous supercurrents of spin-orbit coupled electrons in superconducting two-dimensional systems and thin films. In this work it is shown that current vortices can be created at the ends of a long homogeneously magnetized strip of a ferromagnetic insulator, which is deposited on the surface of a three-dimensional topological insulator. The s-wave superconductivity on its surface is assumed to have an intrinsic origin, or to be induced by the proximity effect. It is shown that vortices with the odd vorticity can localize Majorana zero modes.The latter may  also be induced by sufficiently narrow domain walls inside the strip, that opens a way for manipulating these modes by moving the walls. It is shown that the vorticity can be tuned by varying the magnetization and width of the strip. A stability of the strip magnetization with respect to the Berezinsky-Kosterlitz-Thouless transition has been analyzed.
\end{abstract}
\maketitle

\section{Introduction}

The effect of the Zeeman interaction on the formation of a nonuniform superconducting state is widely studied since the Fulde-Ferrel-Larkin-Ovchinnikov  seminal discovery \cite{Larkin,Fulde} that the Zeeman splitting of Cooper pair electrons  results in the superconducting order parameter which varies periodically in space. A new insight into this research field was brought about by understanding of the role played by the spin-orbit coupling (SOC) of electrons. Edelstein \cite{Edelstein} has shown that the interplay of the Rashba SOC and the Zeeman interactions lead to a spontaneous supercurrent in a two-dimensional (2D) superconductor, if the Zeeman field is parallel to the 2D system. In systems with a homogeneous Zeeman field this effect results in a helix spatial structure of the  order parameter \cite{Edelstein,Samokhin,Barzykin,Agterberg,Kaur,Agterberg2,Dimitrova}, so that the supercurrent turns to zero due to a compensating current which originates from the order-parameter phase gradient. The situation is quite different in systems with a nonuniform Zeeman field, for example, when it is finite only within some regions, which may be created by a proximity of a superconductor to magnetic materials. It was shown that a nonuniform parallel field may induce supercurrents \cite{Malsh island, Pershoguba,Hals} around magnetic islands on the surface of a two-dimensional superconductor. Another group of systems, where this sort of  magnetoelectric effect may be observed, comprises so called  $\phi_0$ Josephson junctions \cite{Krive,Reinoso,Zazunov,ISHE,Liu,Yokoyama,Konschelle,Assouline}. A role of a weak link in these junctions is played by a 2D normal metal, where both SOC and a parallel Zeeman are presented.

In a thermodynamically equilibrium system spacial variations of the order parameter and  supercurrents, which are induced by a nonuniform Zeeman field,  are determined by the energy minimum of the electronic system. In earlier studies only topologically trivial spacial variations of the order-parameter phase have been taken into account. On the other hand, the energy minimum may be reached in a superconducting state which involves supercurrent vortices where this phase winds up integral multiples of $2\pi$ around singular points. Depending on the geometry of a magnetic island these vortices might partly, or completely, compensate the current induced by the Zeeman interaction, and thus reduce the energy of the state. This situation was not addressed yet.

In this work, we just focus on one important application of this idea and consider an example that could serve as a new platform for the localization of  Majorana zero modes (MZM). MZM are localized  quasiparticles whose energy is pinned to the middle of the superconducting gap. These particles have an unusual non-Abelian statistics which, in combination with their resilience with respect to external perturbations, makes them a promising tool for quantum computing \cite{Kitaev,Nayak,Alicea}.  Many efforts have been made to find an appropriate system where this idea may be implemented  \cite{Oreg,Lutchyn,Zhang,Fu,Nadj,Pientka,Christensen,Klinovaya,Menard,Pawlak,Li}. Some theoretical studies predicted that MZM may be localized near vortices in topological superconductors \cite{Volovik,Kopnin,Gurarie,Tewari,Galitski,Jackiw,Read,Ivanov,Sau,Santos}, as well as in three dimensional (3D) topological insulators (TI), where the topological superconductivity  is induced by the proximity effect of an adjacent s-wave superconductor \cite{Galitski,Fu}.

Based on these results it is natural to explore conditions for the formation of MZM-carrying vortices by relying on the combined magnetoelectric effect of SOC and the Zeeman field.  It is shown below that such vortices appear in a 2D system where the Zeeman field has a form of a long and narrow strip whose magnetization  is directed parallel to the strip. Majorana zero modes are expected to localize at its ends, or at domain walls (DW), as shown in Fig.1. It is assumed that superconductivity of Dirac electrons on the TI surface can be induced either by a thin s-wave superconducting film, or may have an intrinsic origin. The Zeeman field, in turn, can be produced by the exchange interaction of electrons with an adjacent magnetic wire. A significant advantage of such MZM's is that the vortices can be created near well defined positions. Moreover, those which are localized at DW may be  manipulated by various methods which are used to move DW.

\section{Model}

Let us consider a system where the width $w$ of the magnetic strip is much less than the coherence length $\xi$ of the superconductor's order parameter $\Delta(\mathbf{r})$. At the same time, the strip length $L$ is large enough, so that $L \gg \xi \gg w$. The small width of the strip is assumed only for the sake of simplicity, because in this limit an analytical solution is possible. The  Hamiltonian of a 2D electron gas on the TI surface is given by $H=\sum_{\mathbf{k}}\psi^{\dag}_{\mathbf{k}}\mathcal{H}\psi_{\mathbf{k}}$, where $\psi_{\mathbf{k}}$ are the electron field operators,  which are defined in the Nambu basis as $\psi=(\psi_{\uparrow},\psi_{\downarrow},\psi^{\dag}_{\downarrow},-\psi^{\dag}_{\uparrow})$ and the one-particle Hamiltonian
$\mathcal{H}$ is given by
\begin{eqnarray}\label{H}
&&\mathcal{H}=v\sigma^x(\tau_3k_y-F_y(\mathbf{r}))-v\sigma^y(\tau_3k_x-F_x(\mathbf{r})) -\nonumber\\
&&\tau_3\mu +Re[\Delta(\mathbf{r})]\tau_1-Im[\Delta(\mathbf{r})]\tau_2\,,
\end{eqnarray}
 where $\mu$ is the  chemical potential,  $F_x=Z_y/v, F_y=-Z_x/v$, $\mathbf{Z}$ is the Zeeman field produced by the exchange interaction of conduction electrons with spins of the magnetic strip, $\mathbf{k}=-i\partial/\partial\mathbf{r}$, and $\sigma^j$ denote Pauli matrices ($j=x,y,z$). The Pauli matrices $\tau_i$, $i=1,2,3$, operate in the Nambu space. It is assumed that $\mathbf{Z}$ is parallel to the $x$-axis which is directed along the strip. Hence, $F_x=0$ and $F_y=-Z_x/v \equiv-Z/v$.
\begin{figure}[tp]
\includegraphics[width=9cm]{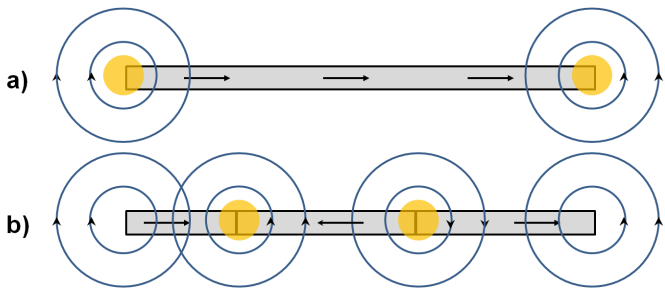}
\caption{(Color online)  a) A Zeeman  field which is induced on the surface is shown as a strip. The magnetization direction is depicted by arrows. MZM and supercurrent vortices are localized near edges of the strip. b) Two MZM are localized near Ising domain walls } \label{fig1}
\end{figure}

It is seen that $\mathbf{F}$ in (\ref{H}) plays the role of a gauge field.  It may induce a supercurrent, similar to the electromagnetic vector-potential.
The current conservation, however, can be guaranteed  only if $\bm{\nabla}\mathbf{F}=0$. For arbitrary $\mathbf{F}$ it can not be reached, because $\mathbf{F}$ is not a true gauge field. Therefore, it can not be modified by the gauge transformation. On the other hand, if $\Delta$ has a varying in space phase, such that $\Delta=|\Delta|\exp( i\theta)$, the supercurrent in a disordered system is given by  $\pi|\Delta| N_F e^2D(\bm{\nabla}\theta - 2\mathbf{F})$, where $N_F$ is the 2D density of electronic states at the Fermi energy and $D$ is the diffusion constant. \cite{Kopnin2} This expression is valid for a Dirac system, as well.  \cite{Zyuzin, Bobkova} From this expression it follows that the current is conserved, if the phase satisfies the equation
\begin{equation}\label{theta}
\nabla^2\theta=2\bm{\nabla}\mathbf{F}
\end{equation}
To clarify the special choice of the Zeeman field in Fig.1, let us consider a limiting case of a very long strip with $L\rightarrow \infty$. By performing a unitary transformation of Eq.(\ref{H}) in the form $U^{\dag}\cal{H}U$, where $U=\exp(i\tau_3\theta/2))$, we arrive at the transformed Hamiltonian $\tilde{\cal{H}}$, which at $Z_y=0$ ($F_y = -Z/v$) has the form
\begin{eqnarray}\label{tildeH}
&&\tilde{\mathcal{H}}=v\sigma^x\left(\tau_3k_y +\frac{\nabla_y\theta}{2}-F_y(\mathbf{r})\right)-\nonumber\\
&&v\sigma^y\left(\tau_3k_x +\frac{\nabla_x\theta}{2}\right)-\tau_3\mu +|\Delta(\mathbf{r})|\tau_1\,.
\end{eqnarray}
It is easy to see that the solution of Eq.(\ref{theta}) is such that $\theta$ is constant outside the strip, where $\nabla_y\theta=0$, and $\nabla_y\theta-2F_y=0$ inside the strip, while $\nabla_x\theta=0$ everywhere. Hence, both the phase and the Zeeman field are removed from  Eq.(\ref{tildeH}). There is no supercurrent and the only effect of the Zeeman field is a linear variation of the order parameter phase inside the strip. This situation differs a lot from conventional superconductors, where at $Z \gg |\Delta|$ the Cooper pairs could penetrate under the strip only within a range much shorter than the superconducting coherence length. The strong spin-momentum locking in the considered Dirac system plays, therefore, a crucial role in diminishing the destructive effect of the Zeeman field. It should be taken into account, however, that  in TI Eq.(\ref{H}) is valid only near the Dirac point. Therefore, $Z$ must be at least much less than the TI insulating gap.

 It is instructive to compare this situation with the case when the strip magnetization is parallel to the $y$-axis. Then, $F_y=0$ and $F_x=Z/v$. In this case the right-hand side of Eq.(\ref{theta}) vanishes and the solution of this equation is $\theta=0$. Therefore, the electric current is absent outside the strip. However, due to the magnetoelectric effect it is fin ite inside it, where $Z\neq 0$. The current  is directed along the $x$-axis and at the small temperature $k_BT \ll |\Delta|$ the corresponding current density is $J_x=eZ\mu/4\pi v$. \cite{Yip}In this case the Zeeman field is not removed from Hamiltonian Eq.(\ref{tildeH}) and produces at $Z\gg |\Delta|$ a destructive effect on the superconductivity and the proximity effect. \cite{Bergeret} Such an interplay between the spontaneous supercurrent and the direction of the strip magnetization  has been previously discussed \cite{Malsh island} for a weakly spin-orbit coupled superconductor.

 Although in the case of $\mathbf{Z}$  parallel to the $x$-axis the Zeeman field can be gauged out far from the strip edges, it retains finite near them. In fact, Eq.(\ref{theta}) describes a 2D capacitor, whose "electric charges" are  accumulated on the lines $y=\pm w/2$, while $\theta$ represents the "electric potential". Accordingly, the current which is proportional to $\bm{\nabla}\theta$ outside the strip, circulates around each edge and decreases as $1/r$ at $r \gg w$, where $r$ is the distance from the edge.  The exponential screening of vortices by an induced magnetic field is absent in the considered 2D case. Note, that in the case of the proximity induced superconductivity the parent superconductor is assumed to be represented by a thin film, which also can not efficiently screen  the vortex. As known \cite{Pearl}, the screening length in a thin film with the thickness $d$  is given by $l_s \equiv l_L^2/d \gg l_L$, where $l_L$ is the London penetration depth. Therefore, as long as  $L\ll L_s$ one may ignore the magnetic screening effect.

 \section{Vortex energy}

 Further, let us consider a situation when a supercurrent vortex is localized near one of the edges, so that the total current  circulating around the edge is a sum of the magnetoelectric current and the current  which is induced by the vortex. By considering the physics near one of the edges we place the $x=0$ point at the edge, while the strip occupies the region $x>0$.  Due to the  long-range $1/r$ decreasing of the current, the most of the vortex energy is accumulated outside the vortex core, at $r \gg \xi$. In this region $|\Delta(\mathbf{r})|\equiv \Delta_0$ is uniform in the space. Therefore, in general the order parameter has the form $\Delta=\Delta_0\exp(i\theta+il\phi)$, where $\phi$ is the polar angle and $l$ is an integer.     It follows from Eq.(\ref{theta}) that $\theta$ is a periodic function of $\phi$. At $r\gg w$ it has the form $-2Zr\phi/v$ at $0<\phi<\phi_0$, $\theta=Zw(\phi-\pi)/\pi v$ at  $\phi_0<\phi<2\pi - \phi_0$, and  $2Zr(2\pi-\phi)/v$, at $2\pi-\phi_0<\phi<2\pi$, where $\phi_0=\arctan(w/2r)$. Hence, at large $x$ the phase $\theta$ winds up from 0 to $2Zw/ v$ around the edge and then, within a narrow angular interval where $\phi$ crosses the strip, returns to its original value. The detailed calculation of the coordinate dependence of $\theta$ near one of the edges is presented in Appendix A. The vorticity $l$ should be calculated by minimizing the free energy. At $r\gg \xi$ the order parameter varies slowly, that allows using the Ginzburg-Landau (GL) formalism for the calculation of the energy $F_v$, which, except for constant terms, is given by
 \begin{eqnarray}\label{Fv}
 &&F_v=\alpha\Delta_0^2\int_{\xi}^L rdrd\phi|(-i \bm\nabla-2\mathbf{F})e^{(i\theta+il\phi)}|^2=\nonumber \\
 &&2\pi\alpha\Delta_0^2\int_{\xi}^L \frac{dr}{r}(\Phi-l)^2=2\pi\alpha\Delta_0^2(\Phi-l)^2\ln\frac{L}{\xi}\,,
 \end{eqnarray}
 where $\Phi=Zw/\pi v$ and $\alpha$ is the Ginzburg-Landau parameter. A relatively small contribution of the core region is excluded from the vortex energy  and enters as a cutoff of the logarithm. The  vorticity $l$ should be calculated by minimizing $(\Phi-l)^2$. The situation resembles the Little-Parks effect \cite{Little} where instead of the external magnetic field flux we have the "Zeeman flux" $\Phi$. Although, physically $\Phi$ has nothing common with the magnetic field flux, in this work it will be called "flux", due to the mentioned formal coincidence. At small $\Phi$ the minimum energy is obtained at $l=0$. At $\Phi=1/2$ there are two degenerate states with $l=1$ and $l=0$. The vorticity $l=1$ is realized at $1/2<\Phi<3/2$. On the opposite end of the strip $\Phi$ changes the sign. Accordingly, $l=-1$, if $-3/2<\Phi<-1/2$. In order to evaluate $\Phi$, let us take $v=$3eV{\AA} in 3D TI  \cite{Qi RMP} and $w=50$nm. Then,  $\Phi=1/2$ may be reached at $Z=10$meV. Therefore, the vortex becomes energetically favorable at moderate values of $Z$.\cite{Z}

 The discussed situation, however, is related to an intrinsic superconductor. In the case of the proximity induced superconductivity one should take into account that the Zeeman flux $\Phi$ is generated in the 2D Dirac system, but not in the 3D proximized  superconductor. At the same time, a vortex produces the current in the whole system. Therefore, $\Phi$ enters in Eq.(\ref{Fv}) with a weight which reduces the efficiency of the Zeeman flux. As an example, let us consider a simple bilayer system where the TI surface and a ferromagnetic insulator wire are buried under a very thin film of an s-wave superconductor. If the thickness $d$ of the film  is much less than the coherence length, the order parameter is uniform throughout the film (in the $z$-direction). A simple, though not rigorous evaluation of the effective $\Phi$ in Eq.(\ref{Fv}) may be obtained by assuming a good contact between the film and the TI surface. In this case the Cooper pair wave function  takes in 2D electron gas the same value as in the film. Hence, the GL functional can be averaged over $z$. By taking into account that the parameter $\alpha$ in Eq.(\ref{Fv}) is proportional to the normal state conductivity \cite{Kopnin2}, one may evaluate the effective flux as  $\Phi_{eff}=b(\sigma_{2D}/\sigma_{s}d)\Phi$, where $\sigma_{s}d$ and $\sigma_{2D}$ are the sheet conductivity of the superconductor film and 2D conductivity of surface electrons, respectively. $b\sim 1$ is a material dependent constant. For  $d=10$nm the typical ratio $\Phi_{eff}/\Phi$ varies between 10$^{-2}$ and 10$^{-3}$. Therefore, it is necessary to have weakly conducting and very thin superconducting films.  At the same time, the employed theoretical approach requires $w \ll \xi$, that does not allow to increase $\Phi$ by using more wide strips. Also, the magnitude of the Zeeman interaction is restricted by the employed model. \cite{Z} On the other hand, recent experimental results, such as  the self-epitaxy induced interface superconductivity on the TI surface,\cite{Bai} go beyond the considered bilayer model. Furthermore, recent measurements \cite{Trang} of Pb films, which were epitaxy grown on TlBiSe$_2$ substrate, demonstrated that the Dirac surface state of TI migrates on top of the Pb film and that this state acquires a superconducting energy gap. In this case the above analysis for an "intrinsic" superconductor becomes valid.

\section{Majorana zero modes}

Now let us consider the Bogolubov-de Gennes (BdG) equations by starting from Eq.(\ref{tildeH}), where the order parameter has the vorticity $l$. Hence, instead of a real and positive $\Delta$ in Eq.(\ref{tildeH}) we have $\Delta=\Delta_0 f(\mathbf{r})\exp(il\phi)$, where the dimensionless function $\mathbf{r}$ describes the vortex core, so that $f(r)\rightarrow 0$ at $r\ll \xi$ and $f(r)\rightarrow 1$ at $r\gg \xi$. It is important that a detailed knowledge of the  order parameter behavior within the vortex core is not important for the calculation of the  MZM wave function \cite{Galitski}.

Note that the function $\mathbf{A}\equiv\bm{\nabla}\theta/2 - \mathbf{F}$, which appears in Eq.(\ref{tildeH}), satisfies the equation $div\mathbf{A}=0$. Therefore, it can be represented as $\mathbf{A}=curl \mathbf{B}$, where the vector $\mathbf{B}$ is perpendicular to the surface of TI. By calculating  $curl\mathbf{A}$  one obtains
\begin{equation}\label{A}
curl\mathbf{A}=-curl\mathbf{F}=-\nabla^2 \mathbf{B}\,.
\end{equation}
By substituting $F_y=-Z/v$, $F_x=0$ we get near an isolated edge of the strip $(curl\mathbf{F})_z=\partial Z/\partial x=(Z/v)\delta(x)\eta(w/2-y)\eta(w/2+y)\equiv 2\pi\rho$, where $\eta(x)$ denotes the Heaviside step function. Hence $B_z\equiv B$ satisfies the Poisson equation
\begin{equation}\label{B}
\nabla^2 B=2\pi\rho(\mathbf{r})
\end{equation}
with the "charge" density $\rho(\mathbf{r})$ distributed exactly on the short edge of the rectangular strip. The edge may be more smooth when $Z$ gradually decreases to zero outside the strip, that is more realistic from the experimental point of view. In this case $\rho$ will be distributed near the edge over a region with the size $\sim w$. At the large distance $r\gg w$ the solution of Eq.(\ref{B}) has the form $B=(\Phi/2) \ln r$, where $\Phi=2\int \rho d^2 r $. Since the MZM is expected to localize within the distance $\sim \xi \gg w$, it is reasonable to use this logarithmic form of $B$ in BdG equations. For simplicity it will be assumed below that $\Phi>0$.  In case of MZM we look for a nondegenerate eigenstate of the BdG equation which has the zero energy. The corresponding wave function satisfies the equation $\tilde{H}\Psi(\mathbf{r})=0$, where the four-vector $\Psi$ has the components $u_{\uparrow},u_{\downarrow},v_{\downarrow}$ and $-v_{\uparrow}$. Here, the arrows denote the spin projection, while $u$ and $v$ denote variables in the Nambu space. The particle-hole symmetry requires MZM to be an eigenstate of the charge conjugation operator $\tau_2\sigma^y K$, where $K$ is the complex conjugation. Therefore, $u_{\uparrow}=v^*_{\uparrow}$ and $u_{\downarrow}=v^*_{\downarrow}$. The following analysis of BdG equations is based on previous calculations of MZM localized on a vortex. According to \cite{Galitski}, the MZM solution of BdG equations may be obtained at odd $l$. The corresponding wave function has the form $u_{\uparrow}=\exp[i(\frac{l-1}{2})\phi-i\frac{\pi}{4}]\chi_{\uparrow}$ and $u_{\downarrow}=\exp[i(\frac{l+1}{2}))\phi+i\frac{\pi}{4}]\chi_{\downarrow}$. Other components of the Nambu spinor can be obtained from the above charge conjugation relations.  The functions $\chi$ are real and satisfy the equation
\begin{equation}\label{bdg}
\left( {\begin{array}{*{20}{c}}
{ - \mu }&M_{\uparrow\downarrow}\\
M_{\downarrow\uparrow}&{ - \mu }
\end{array}} \right)\left( {\begin{array}{*{20}{c}}
\chi_{\uparrow} \\
\chi_{\downarrow}
\end{array}} \right)=0\,,
\end{equation}
where in polar coordinates the matrix elements $M$ can be written as
\begin{eqnarray}\label{M}
M_{\uparrow\downarrow}=v\left(\nabla_r+\frac{l+1}{2r}-\frac{\Phi}{2r}\right)+\Delta_0f \,,\nonumber \\
M_{\downarrow\uparrow}=-v\left(\nabla_r-\frac{l-1}{2r}+\frac{\Phi}{2r}\right)-\Delta_0f\,.
\end{eqnarray}
Unlike the previously studied BdG equations, here the new term $\Phi/2r=\nabla_r B$ appears in Eq.(\ref{M}), where the long-range form of $B$, namely, $B=(\Phi /2)\ln r$ is taken.  Up to a normalization factor, the functions $\chi$ are obtained from Eq.(\ref{bdg}) in the form
 \begin{equation}\label{psi}
\left( {\begin{array}{*{20}{c}}
\chi_{\uparrow} \\
\chi_{\downarrow}
\end{array}} \right)=
\left( {\begin{array}{*{20}{c}}
J_{ |l-\Phi-1|/2}(\mu  r/v)\\
J_{ |l-\Phi-1|/2+1}(\mu r/v)
\end{array}} \right)e^{-\Delta_0\int_0^r dr^{\prime}f(r)/v}\,,
\end{equation}
where $J_{ \nu}(\mu  r/v)$ is the Bessel function. Note, that this expression for the MZM wave-function is valid only at the large distance $r\gg w$ from the edge. At distances $r\lesssim w$ Eq.(\ref{M}) includes additional terms from $B$ which decrease at large $r$ faster than $\Phi/r$. The effect of these terms can be easy analyzed in the case when the "flux" density $\rho$ in Eq.(\ref{B}) is isotropic due to smooth edges of the wire. It can be shown (see Supplementary Material) that the effect of the near field is only to modify the spinor $\chi$ in Eq.(\ref{psi}).

\section{Majorana zero modes localized at domain walls}

A vortex can be induced by DW. Let us consider a simple case of the Ising wall shown in Fig.1c. The strength of such a vortex is determined by the integrated Zeeman flux density $\rho(\mathbf{r})$ in  Eq.(\ref{B}). This strength $\Phi_{dw}$ is twice the strength of the vortex near the edge of the wire. Hence, the above analysis may be applied to the case of a DW, with the substitution $\Phi_{dw} = 2\Phi$. As shown above, near a single edge the vortex with $l=1$ appears  if $1/2 <\Phi < 3/2$. Therefore, a domain wall may localize the $l=1$ vortex at $1/4 <\Phi < 3/4$.  A magnetic wire can carry several DWs.  At the same time, the total Zeeman flux including DWs and edges is zero. It can be seen by integrating Eq.(\ref{A}) over the region enclosing the entire strip. The corresponding contour integration of $\mathbf{A}$ gives zero, because $\theta$ is a periodic function  and  $\mathbf{F}=0$ outside the strip. This sum rule imposes a restriction on the total vorticity $l_{\mathrm{tot}}=\sum_i l_i$, where $l_i$ is the vorticity of $i$-th  vortice, including DW and edges. It follows from Eq.(\ref{Fv}) where, at large distances, $\Phi=0$ (the total Zeeman flux) and $l=l_{\mathrm{tot}}$. At the finite $l_{\mathrm{tot}}$ the upper cutoff of the integral is determined by the size of the system $\rightarrow \infty$ or by the large magnetic screening length $l_s$ in thin films. Therefore, $l_{\mathrm{tot}}$ must be zero. Otherwise, we get the infinite energy. If the number of DW is even, the sum rule is satisfied automatically, because in this case the fluxes and vorticities at the ends of the wire have opposite signs and their sum is zero. The same takes place for domain walls, since there is the equal number of  the walls with "positive" and "negative" fluxes. In contrast, in the case of the odd number of DW  the edges carry fluxes of the same sign. Therefore, their total vorticity is even. It can be compensated only by the same vorticity carried by DW. For example, if there is a single DW, its vorticity should be even. But such DW can not localize MZM. Therefore, MZMs can not reside on a single and, pressumably, any odd number of  DWs, if the wire's width is uniform, as in Fig.1.

The above analysis  has been restricted to a simple case of Ising DWs. On the other hand, rich opportunities to manipulate DW arise in the case of the Bloch or Neel DW. Their study is outside the scope of this work. It is reasonable, however, to assume that if the size of these walls is much less than $\xi$, an internal structure of DW is not important.

\section{Stability of the strip}

Domain walls may be spontaneously created  by pairs in a wire. Depending on the temperature in a thin wire there is some number of thermally excited DW. However, the $\ln L$ dependence of their energy, which is associated with vortices, can lead to the Berezinskii-Kosterlitz-Thouless (BKT) \cite{Berezinskii,Thouless} transition at  the temperature $T>T_{\mathrm{BKT}}$. $T_{\mathrm{BKT}}$ can be evaluated from the  energy  Eq.(\ref{Fv}) of a single vortex, which resides on a DW. In this case, according to Refs.[\onlinecite{Berezinskii,Thouless}] the BKT transition temperature is given by
\begin{equation}\label{BKT}
k_B T_{\mathrm{BKT}}=\pi\alpha\Delta_0^2(\Phi-l)^2\,.
\end{equation}
By taking into account that $\alpha \simeq \xi^2 N_F$, one obtains $k_B T =E(\Phi-l)^2$, where the energy $E$ is determined by parameters of either an intrinsic 2D superconductor, or parameters of a 3D superconducting film, in case of proximity induced superconductivity. Anyway, $E$ is of the order of electronvolts, or in the case of a dirty metal may be tenths of 1eV. Therefore, at $(\Phi-l)^2\sim 1$ the corresponding $T_{\mathrm{BKT}}$ is much larger than the temperature range of interest. A special case is  $(\Phi-l)^2\rightarrow 0$. However, one has to take into account a quite large (for an Ising DW) exchange energy stored in DW. It may be ignored only at a very large length of the wire, which is always restricted by experimental conditions.

\section{Conclusion}

In conclusion, it is shown that supercurrent vortices can be induced by a Zeeman field at the ends of a long ($L\gg \xi$) ferromagnetic insulator wire which is deposited on the superconducting surface of a 3D topological insulator, provided the magnetization is parallel to the wire's long side. In such a geometry even a strong exchange interaction of 2D Dirac electrons with spins of the ferromagnetic insulator can destroy superconductivity (or the proximity effect) only near the ends of the wire, at $r \ll \xi$. Conditions for the localization of MSM at these vortices have been considered. Within a reasonable range of parameters the odd vorticity can be achieved, that is necessary for the MZM formation. Similar situation takes place near a domain wall. Only a simple case of the Ising DW was considered. It was assumed that the superconductivity of 2D Dirac electrons is induced by the proximity effect of an s-wave superconductor. A simple evaluation for a bilayer system which consists of a thin superconducting film deposited at the surface of TI, has shown that it is difficult to reach the high enough strength of the effective Zeeman flux, at least within limitations of the chosen model and approximations of the theory. At the same time, there can be quite different scenarios for the proximity effect. For example, the Dirac surface state of a TI may migrate on top of a very thin superconducting film, instead of staying buried under it. Moreover, this Dirac band acquires a superconducting gap, as was observed in Ref.[\onlinecite{Trang}]. In this case the model of, what was called above, "intrinsic" 2D superconductor might be applicable.

\emph{Acknowledgements} - The work was partly supported by the Russian Academy of Sciences program "Actual
 problems of low-temperature physics."
%%%%%%%%%%%%%%%%%%%%%%%%%%%%%%%%%%%%%%%%%%%%%%%%%%%%%%%%%%%%%%%%%%%%%%
%%%%%%%%%%%%%%%%%%%%%%%%%%%%%%%%%%%%%%%%%%%%%%%%%%%%%%%%%%%%%%%%%%%%%%%

%%%%%%%%%%%%%%%%%%%%%%%%%%%%%%%%%%%%%%%%%%%%%%%%%%%%%%%%%%%%%%%%%%%%%%%%
%%%%%%%%%%%%%%%%%%%%%%%%%%%%%%%%%%%%%%%%%%%%%%%%%%%%%%%%%%%%%%%%%%%%%%%
\appendix

\section{The phase of the order parameter and the  supercurrent induced by the Zeeman interaction}

Let the Zeeman field $\mathbf{Z}$ to be directed parallel to the $x$-axis. This field is nonzero in a strip which occupies the region: $0<x<L$ and $-w/2<y<w/2$. It will be assumed that $L\gg w$. The phase, which is induced by $Z$ satisfies  Eq.(\ref{theta}). For the chosen geometry of the ferromagnetic strip the solution of this equation has the form
\begin{eqnarray}\label{thetasolve}
&&\theta(x,y)=-F_y\int_{0}^{L}\frac{dx^{\prime}}{2\pi}\left[\ln\left((x-x^{\prime})^2+(y-\frac{w}{2})^2\right)-\right.\nonumber\\
&&\left.\ln\left((x-x^{\prime})^2+(y+\frac{w}{2})^2\right)\right]\,.
\end{eqnarray}
In the range of $x\ll L$ and $y\ll L$ the integration may be extended to $L\rightarrow \infty$. This results in
\begin{widetext}
\begin{eqnarray}\label{thetasolve2}
&&\theta(x,y)=-\frac{F_y}{2\pi}\left[\left(r^{+}-i\frac{w}{2}\right)\ln\left(r^{+}-i\frac{w}{2}\right)+\left(r^{-}+i\frac{w}{2}\right)\ln\left(r^{-}+i\frac{w}{2}\right)- \right.\nonumber \\
&& \left. \left(r^{+}+i\frac{w}{2}\right)\ln\left(r^{+}+i\frac{w}{2}\right)-\left(r^{-}-i\frac{w}{2}\right)\ln\left(r^{-}-i\frac{w}{2}\right)\right]\,,
\end{eqnarray}
\end{widetext}
where $r^{+}=x+iy$ and $r^{-}=x-iy$. The induced supercurrent is proportional to $\bm{\nabla}\theta - 2\mathbf{F}$. A discontinuity of $\bm{\nabla}\theta$ on the strip boundary is just to compensate $2\mathbf{F}$ in the latter equation. Therefore, $\mathbf{A}=(\bm{\nabla}\theta/2) - \mathbf{F}$ is a regular function, with the exception of singularities at points with the coordinates $x=0,y=w/2$ and $x=0,y=-w/2$. A similar behavior takes place near the opposite end of the strip at $x=L$.

By expressing $F_y$ as $-Z/v$, in the leading approximation with respect to $w/r$ one can obtain from Eq.(\ref{thetasolve2})
\begin{eqnarray}\label{thetasolve3}
\theta(x,y)&=&\frac{Zw}{\pi v}\phi\,\, \,\,\,\,\,\, \left(\arctan\frac{w}{2y}<\phi <  2\pi -\arctan\frac{w}{2y}\right)\,,\nonumber \\
\theta(x,y)&=&\frac{Zw}{\pi v}\phi-r\frac{2Z}{v}\phi\,\, \,\,\,\,\,\, \left(|\phi| < \arctan\frac{w}{2y}\right)\,,
\end{eqnarray}
where $\phi$ is the polar angle.

The  "vector potential" $\mathbf{A}$ can be written in the form of an expansion in powers of $w/r$. By expanding Eq.(\ref{thetasolve2}) one obtains
\begin{eqnarray}\label{thetasolve}
A_x=-\frac{Z\gamma}{\pi v}\left(\sin\phi + \sum_1^{\infty}\frac{(-1)^{k}\gamma^{2k}}{2k+1}\sin(2k+1)\phi\right) \nonumber\\
A_y=\frac{Z\gamma}{\pi v}\left(\cos\phi + \sum_1^{\infty}\frac{(-1)^{k}\gamma^{2k}\cos(2k+1)\phi}{2k+1}\right)\,,
\end{eqnarray}
where $\gamma=w/2r$. The leading term in $\mathbf{A}$ coincides with the electromagnetic vector potential produced by the magnetic flux $Zw/2\pi v$ piercing the $r=0$ point. Accordingly, the induced supercurrent is proportional to $\mathbf{A}$.

\section{The Majorana zero mode localized near a smooth edge of a Zeeman strip}

As shown in the main text, the vector $\mathbf{A}$ may be represented as $\mathbf{A}=\bm{\nabla}\times\mathbf{B}$, where $\mathbf{B}$ is perpendicular to the $xy$ plane and $B_z\equiv B$. It is reasonable to consider a situation when the source $\rho$ in  Eq.(\ref{B}) is isotropically distributed around the point $r=0$. It is assumed that $\rho$ decreases fast at $r>w/2$. The Zeeman field $Z_x(\mathbf{r})$, which produces  $\rho(r)$ of a given form, is expressed as $Z_x(\mathbf{r})=2\pi\int_{\infty}^x dx^{\prime} \rho(r^{\prime})$. By expressing $\mathbf{A}$ in BdG equations  in terms of $B_z$ one can write Eq.(\ref{M}) in the form
\begin{eqnarray}\label{M2}
M_{\uparrow\downarrow}=v\left(\nabla_r+\frac{l+1}{2r}-\frac{\partial B}{\partial r}\right)+\Delta_0f \,,\nonumber \\
M_{\downarrow\uparrow}=-v\left(\nabla_r-\frac{l-1}{2r}+\frac{\partial B}{\partial r}\right)-\Delta_0f\,.
\end{eqnarray}
By substituting these matrix elements into Eq.(\ref{bdg}) we obtain the following equation for $\tilde{\chi}_{\uparrow}=\chi_{\uparrow}\exp(\Delta_0\int_0^r dr^{\prime}f(r)/v)$:
\begin{equation}\label{chup}
-\frac{\partial^2\tilde{\chi}_{\uparrow}}{\partial r^2}-\frac{1}{r}\frac{\partial\tilde{\chi}_{\uparrow}}{\partial r}+\left(\frac{\partial B}{\partial r}-\frac{l-1}{2r}\right)^2\tilde{\chi}_{\uparrow}-2\pi\rho(r)\tilde{\chi}_{\uparrow}=\frac{\mu^2}{v^2}\tilde{\chi}_{\uparrow}\,.
\end{equation}
In turn, the function $\chi_{\downarrow}$ can be expressed in terms of $\chi_{\uparrow}$ by using Eq.(\ref{bdg}). It is easy to see that by  ignoring the short range terms in Eq.(\ref{chup}) at $r\gg w$  this equation can be reduced to the Bessel equation which gives the solution  Eq.(\ref{psi}). In general, Eq.(\ref{chup}) looks as a Schr\"{o}dinger equation for a particle with the positive "energy" $\mu^2/v^2$ and the angular moment $(l-1)/2$, which moves in a cylindrically symmetric potential.
At $l=1$ we have a repulsive and attractive parts of the potential, depending on the sign of $\rho(r)$. Both parts are regular functions at $r=0$. From the general properties of the Schr\"{o}dinger equation it is clear that  the short-range potential will modify the  asymptotic behavior by producing an additional  phase shift of the MZM wave function. The calculation of the concrete behavior of the wave function is out of the scope of the present work.

\begin{thebibliography}{99}
\bibitem{Z}
The 2D Hamiltonian in this work is a low-energy expansion near the Dirac point. Therefore, $Z$ must  at least be much less than the insulating gap of TI.
\end{thebibliography}
\end{document}